\documentclass[aps,a4paper,showpacs,amsmath,amsfonts,amssymb,footinbib,twocolumn]{revtex4}

\usepackage{graphicx}
\usepackage{epsfig}

\def\(({\left(} \def\)){\right)}
\def\[[{\left[} \def\]]{\right]}

\newcommand{\be}{\begin{equation}}
\newcommand{\ee}{\end{equation}}
\newcommand{\bea}{\begin{eqnarray}}
\newcommand{\eea}{\end{eqnarray}}
\newcommand{\e}{\emph {e}}

\renewcommand{\>}{\rangle}

\begin{document}
\date{\today}

\title{Glassy dynamics as a melting process \\ {\it On melting dynamics and the glass transition, Part II} }

\author {Florent Krzakala $^{1,2}$ and Lenka Zdeborov\'a $^{2,3}$}
\affiliation{$^1$ CNRS and ESPCI ParisTech, 10 rue Vauquelin, UMR 7083 Gulliver, Paris 75005 France \\
  $^2$ Theoretical Division and Center for Nonlinear Studies, Los
  Alamos National Laboratory, NM 87545 USA\\
  $^3$  Institut de Physique Th\'eorique, CEA/DSM/IPhT-CNRS/URA 2306
  CEA-Saclay, F-91191 Gif-sur-Yvette, France
}

\begin{abstract}
  There are deep analogies between the melting dynamics in systems
  with a first order phase transition and the dynamics from
  equilibrium in super-cooled liquids. For a class of Ising spin
  models undergoing a first order transition -- namely $p$-spin
  models on the so-called Nishimori line -- it can shown that the
  melting dynamics can be exactly mapped to the equilibrium dynamics.
  In this mapping the dynamical ---or mode-coupling--- glass
  transition corresponds to the spinodal point, while the Kauzmann
  transition corresponds to the first order phase transition itself.
  Both in mean field and finite dimensional models this mapping
  provides an exact realization of the random first order theory
  scenario for the glass transition. The corresponding glassy
  phenomenology can then be understood in the framework of a standard
  first order phase transition.
\end{abstract}


\maketitle

There is still no agreement on what is the fundamental principle
behind the experimentally observed abrupt change in the relaxation
time of super-cooled liquids when the temperature is lowered
\cite{GLASS-GEN:1,GLASS-GEN:2,ANGELL}. Many scenarios and theories
have been proposed over the time to describe the nature of glasses,
and over the last few years the theoretical research has been
concentrated on the following questions. Is there an underlying
critical phenomenon behind the glass transition or not? Is the glass
transition a thermodynamic or a purely dynamic notion? What is the
correct theory of super-cooled liquids?

The most remarkable experimental fact in the phenomenology of super-cooled
liquids is the extremely fast rise of the relaxation time $\tau$, that
increases by several orders of magnitude as the temperature is
decreased by only a few percents. The relaxation time is traditionally
fitted (and well approximated) by the Vogel-Fulcher-Talman law $\tau
\propto \exp{[A/{\((T-T_{VFT}\))}]}$ \cite{VogelFulcher}. The
extrapolated temperature $T_{VFT}$ is found to be close the Kauzmann
temperature $T_K$ where the extrapolated entropy of the super-cooled
liquid becomes smaller than the entropy of the crystal
\cite{Kauzmann}, a fact that has led to speculations on the
existence of an ideal ---yet impossible to observe in finite time---
glass transition at $T_K \approx T_{VFT}$. Another fact pointing in
the direction of an ideal glass transition is the Adam-Gibbs
\cite{AdamGibbs} relation $\tau \propto \exp{[C/\Delta S(T)]}$, where
$\Delta S(T)$ is the difference between the entropy of the crystal and
the liquid, this gives a further link between thermodynamic and
dynamic behaviour.

In the last decade, a lot of attention has been devoted to growing
length scales. In most theories the slower-than-exponential relaxation
time in fragile super-cooled liquids follows from the fact that larger
and larger regions become correlated as the temperature is lowered, so
that larger ensembles of particles have to be rearranged collectively
to relax the system into equilibrium. However, the standard static
correlation function does not show any sign of such a growing
correlation length, and more complex correlation functions thus have
to be considered. Two length scales that are observed to grow
significantly when the temperature is lowered have been now
identified. The first is a purely dynamic length associated to spacial
heterogeneities in the dynamics \cite{Silvio,DynBB,DynHS}. The second
length is an equilibrium one associated with the correlation of a
sub-system with its frozen boundaries
\cite{BiroliBouchaud,DynamicBethe,PointToSet}. The existence of
diverging correlation lengths points towards the glass transition
being a critical phenomena.

Is there really a genuine glass transition? Are the time and length
scales really diverging when approaching the glass transition or are
they only growing? A definitive answer on these questions is difficult
to obtain as both simulations and experiments are faced with the
extremely slow dynamics. According to the random first order theory
(RFOT) of the glass transition \cite{KT,KW1,KW2,KTW,GlassMezardParisi}, the
above mentioned time and length scales have a genuine divergence at
the ideal glass transition temperature, although the RFOT theory is
not free from criticisms (see for instance
\cite{Langer,BiroliBouchaud09} and references therein).

In a companion paper \cite{US-PART-I} we have discussed that a large
part of the glassy phenomenology also appears in the melting process
of a fully ordered phase above an ordinary first order phase
transition. The bottom-line of the analysis is that when the ordered
system is brought at higher temperature than the melting point, it is
inside a metastable state from which it needs to escape, and this is
associated to diverging time and length scales analog to those we
just discussed for glasses, except that in this case the existence of
a genuine transition is doubtless. There are, however, also important
differences between melting above an ordinary first order phase
transition and the glassy dynamics.

In this work we will show that for a class of spin models, both in
mean field and finite dimensional systems, these differences are
washed away and the melting dynamics can be shown to be {\it exactly}
equivalent to the equilibrium dynamics above the first order phase
transition.  These models are nothing but variants of the Ising
$p$-spin models that have inspired the RFOT
theory~\cite{REM,P-SPIN,XORSAT}, and most of our approach is built on
the construction of Nishimori and collaborators
\cite{Nishimori-Original,Nishimori}, in particular Ozeki
\cite{Ozeki1,Ozeki2}. Our results are two-folds: (1) we show that the
standard mean-field approach to the glass transition is {\it exactly}
mappable to a melting problem {\it of some sort}; (2) we show that
there exists a set of finite dimensional models where the melting
process is equivalent to the glassy dynamics in a glass-forming
liquid, and that the existence of a first-order transition for the
melting problem implies the existence of a RFOT-like transition for
the glass.  Our results offer an alternative and potentially fruitful
way of looking at the glass transition problem, all from a
theoretical, numerical and experimental point of view. In particular,
many questions about the glass transition may be recasted into the
more familiar and simpler to describe properties of first order phase
transitions.

The paper is organized as follows: In the section I we present a class
of disordered Ising spin models and concentrate on a special line in
the temperature/disorder plane ---the Nishimori line--- where the
melting problem is equivalent to the equilibrium dynamics.  In section
II we discuss the mean-field version of these models. In section III
we concentrate on a three-dimensional case.  We summarize and discuss
our results in the last section.

\section{When melting is equivalent to equilibrium dynamics}
\label{Nishi}
Following the ideas of Edwards and Anderson, a large part of the
progress in the theory of glasses have originated in studies of spin
glasses \cite{EA,MPV,Hertz}. In particular, the Ising $p$-spin glass
\cite{REM,P-SPIN,XORSAT} provides a mean-field theory for the
structural glass transition \cite{KT,KW1,KW2,KTW}, which is an exact
realization of the early landscape picture of Goldstein
\cite{Goldstein}. Here we shall follow this path, although we will not
restrict ourself to the mean field theory.

Consider the $p$-spin Hamiltonian, with $p=3$ we have
\be 
{\cal H}=-\sum_{ijk} J_{ijk} S_{i}
S_{j} S_{k} ,
\label{PSPIN-H}
\ee 
where the sum is over some triplets of spins (the precise details on
how the triplets are chosen depends on the geometry of the problem:
mean-field lattice, finite-dimensional grid, etc.), and the
interactions $J_{ijk}$ are quenched random variables taken from the
distribution
\be 
P_\rho(J_{ijk})= \rho \delta(J_{ijk}-1) + (1-\rho) \delta(J_{ijk}+1)\, ,
\label{disorder}
\ee
where $\rho$ can vary from $\rho=1/2$ (the spin glass case: $J_{ijk}=\pm 1$
with equal probability) to $\rho=1$ (the ferromagnetic case with
$J_{ijk}=1$).

Consider first the {\it pure} ferromagnetic model with $\rho=1$. The
ferromagnetic many-body interaction models undergo usually a
first order transition, both in mean field \cite{US-PART-I,XORSAT} and
in finite dimension (see the simulations of the {\it plaquette models}
in \cite{Plaquette_L,Plaquette_B,STAR_model}). At high temperature
the system is in a paramagnetic/liquid state, while for low
temperature it is in a ferromagnetic/crystal one, and a first order
transition at $T_F$ separates the two.

The dynamical behavior of the ferromagnetic many-body interaction
models, on the other hand, reproduces many aspects of super-cooled
liquids and their glass transition. In particular, crystallization at
$T_F$ seems to be easily avoidable when cooling down from a large
temperature, and the super-cooled liquid so obtained has all the
desired glassy phenomenology. There are, however, several problems
(actually common to most glassy systems) that prevent
analytical and conceptual progress:

(1) The fact that below the ferromagnetic/melting temperature $T_F$
the true equilibrium state is given by the ferromagnet/crystal makes
always any statement on {\it equilibrium} super-cooled liquid
delicate, to say the least.  Any discussion involving the description
of the putative ideal glass transition at temperature $T_K<T_F$ is
plagued by this problem, since there will always be a temperature
beyond which the nucleation time towards the crystal will be larger
than the relaxation time in the super-cooled liquid phase
\cite{Kauzmann,STAR_model}. It would thus be interesting to
find a model with $T_K=T_F$ so that the divergence of the
equilibration time in the liquid phase could be well defined.

(2) Glass formers are known to be hard to simulate, since the time to
find an equilibrium configuration grows faster than exponentially with
the temperature. It would be really convenient to have an equilibrium
configuration to start the simulation with at all temperatures.

(3) Usually in glassy models, it is really difficult to make any
statement which is not coming from numerical simulations. It would be
really convenient to have some analytical results and guarantees of
genuine phase transitions and divergences.

Quite surprisingly, there is a conceptually simple, and rigorous way,
to avoid the above difficulties for a class of Ising $p$-spin models
if one works on the so-called Nishimori line and this is the main
topic of this paper.

\subsection{Physics on the Nishimori line}

\begin{figure}[t]
\hspace{-1cm}
\includegraphics[width=9cm]{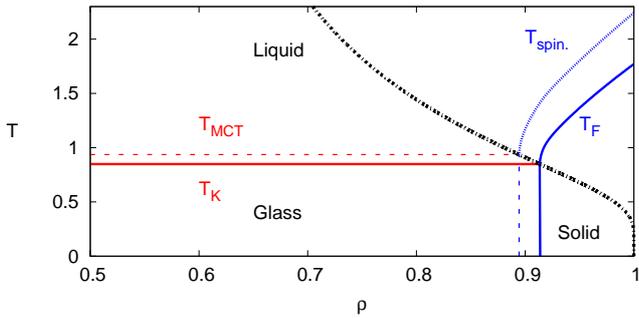}
\caption{(color online) Phase diagram of the mean field $3$-spin Ising
  spin glass model on a Bethe lattice with coordination $c=5$ and a
  fraction of ferromagnetic couplings $\rho$. In the spin glass case
  ($\rho=1/2$) ---which is the starting point of the random first
  order theory of the glass transition--- there is a mode-coupling
  type transition at $T_{\rm MCT}=0.936$ (dashed red) while the
  thermodynamic glass transition is at $T_K= 0.849$ (full red). In the
  purely ferromagnetic case ($\rho=1$) the model undergoes a first
  order ferromagnetic phase transition at $T_F$. The ferromagnetic
  transition extends to lower $\rho$ (lower blue line), as well as the
  spinodal line of the ferromagnetic phase $T_{\rm spin}$ (upper blue
  line). Using a gauge transformation we show that the relaxation
  dynamics for all $T\ge T_K$ in the spin glass case is strictly
  equivalent to the melting dynamics starting from all spins up, $S_i=1$,
  on the Nishimori line, eq.~(\ref{nishimori}), (the black
  line crossing the diagram from upper left to lower right). \label{FigPHASE}}
\end{figure}

As discovered by Nishimori \cite{Nishimori-Original,Nishimori}, there is a
special line, the Nishimori line (NL), in the
temperature-disorder phase diagram where many results can be
established rigorously, on {\it any} lattice and thus in {\it any
  dimension}. Following \cite{Nishimori-Original,Nishimori}, we
start by recognizing that eq.~(\ref{disorder}) can be rewritten as
\be 
P_\rho(J_{ijk})= \frac{e^{K_\rho J_{ijk}}}{2 \cosh  K_\rho} \text{~~~with~~~}
\e^{2K_\rho}=\frac \rho{1-\rho}\, .
\label{nishimori}
\ee 
The Nishimori line in the plane $(T,\rho)$ is defined by
$K_\rho=\beta$ (see Fig.~\ref{FigPHASE} for an explicit example). 

Consider a given quantity of interest $A$, say the free energy, the energy or the magnetization, that depends on the realization of
the disorder $\{{J_{ijk}\}}$, on the inverse temperature $\beta$, and 
in general also
on the spin configuration $\{S_i\}$. As is usual for random systems, we want
to consider the average of this quantity over the disorder
parameterized by the variable $\rho$.
\be 
[A]_{\rho} \equiv \sum_{\{J_{ijk}\}} A(\{ J_{ijk}\},\{S_i\},\beta)  \prod_{ijk} \frac{\e^{K_{\rho} J_{ijk}}}{2 \cosh K_\rho} \, .
\ee
Following again \cite{Nishimori}, we consider a new set of Ising
($\pm 1$) variables $\{{\sigma_i\}}$ and the following gauge
transformation
\bea
S_i &=&\tilde S_i\sigma_i\, , \label{gauge_s}\\
J_{ijk}&=& \tilde J_{ijk} \sigma_i \sigma_j \sigma_k\, . \label{gauge_j}
\eea
It is easy to check that this leaves the Hamiltonian invariant,
i.e. ${\cal H}(\{S_i\},\{J_{ijk}\})={\cal H}(\{S_i\sigma_i\},\{
J_{ijk} \sigma_i\sigma_j\sigma_k\})$. We now apply the gauge
transformation and obtain:
 \bea && [A]_{\rho} = \sum_{\{\tilde
  J_{ijk}\}} \frac{\e^{K_{\rho} \sum_{ijk}\tilde J_{ijk}
    \sigma_i \sigma_j \sigma_k }}{2^M (\cosh{K_{\rho}})^M} \nonumber \\
&& A(\{\tilde J_{ijk} \sigma_i \sigma_j \sigma_k \},\{\tilde
S_i\sigma_i\},\beta)\, , \eea
where $M$ is the number of interactions. Note that the set of values over which is performed the sum $\sum_{J_{ijk}}$ and $\sum_{S_i}$ does not change under the transformation either. We now average quantity $[A]_{\rho}$ over all the $2^N$ possible choices of the gauge transformation $\{\sigma_i\}$ to obtain:
\bea
&&[A]_{\rho} = \frac
{1}  {2^N(\cosh K_{\rho})^{M}} \sum_{\{J_{ijk}\}} \frac 1 {2^M}  \nonumber \\ &&
\sum_{\{\sigma_i\}} \,  \e^{-K_{\rho} {\cal H}(\{\sigma_i\}) }  A(\{J_{ijk} \sigma_i \sigma_j \sigma_k \},\{S_i\sigma_i\},\beta)
 \, .
\label{NL-eq:1}
\eea
At this point, one recognizes that the denominator of the first term is
nothing but the {\it annealed average} of the partition sum for
$\rho=1/2$, and the second term is the disorder average of $A\exp{(-K_\rho {\cal H})}$ for $\rho=1/2$. In other words
\be [A\,]_{\rho} = \frac {\left[ \sum_{\{\sigma_i\}} \,  \e^{-K_\rho
      {\cal H}(\vec{\sigma})} A(\{J_{ijk} \sigma_i \sigma_j \sigma_k \},\{S_i\sigma_i\},\beta)
  \right]_{\frac{1}{2}}}{\left[Z\((K_\rho\))\right]_{\frac{1}{2}}}\, . \label{blabla}
\ee
If we are on the Nishimori line $K_\rho=\beta$, we recognize the
right hand side to be the annealed thermodynamic average of the
quantity $A$.  If $A$ itself is a gauge invariant thermal average of
some quantity then from (\ref{blabla}) follows that on any lattice the
quenched thermodynamic average on the Nishimori line is equal to the
annealed thermodynamic average in the fully disordered case ($\rho=1/2$)\footnote{This was observed in \cite{Antoine} in a particular case of eq.~(\ref{blabla}) when $A=Z$, leading to $[Z^n]_{\rho_{\rm NL}}=[Z^{n+1}]_{1/2}/[Z]_{1/2}$.}.

Let us denote $Z(\{J_{ijk}\},K_\rho)$ the partition function for a given realization of disorder $\rho$ and inverse temperature $K_\rho$. We denote the thermal average with respect to $\{\sigma_i\}$ at inverse temperature $K_\rho$ as 
\be
   \langle A(\{\sigma_i\}) \rangle_{K_\rho} \equiv \frac{ \sum_{\{\sigma_i\}} \, A(\{\sigma_i\}) \,  \e^{-K_\rho
      {\cal H}(\vec{\sigma})}}{Z(\{J_{ijk}\},K_\rho)} \, .
\ee
We now wish to prove the following identity:
\bea &&[A(\{J_{ijk}\},\{S_i\},\beta)]_{\rho} = \nonumber \\
&&\left[\<A(\{J_{ijk} \sigma_i \sigma_j \sigma_k \},\{S_i\sigma_i\},\beta)  \>_{K_{\rho}} \right]_{\rho}\, .
\label{theorem}
\eea
It has a simple interpretation (that will get clearer with the
examples of the next section): consider an instance with disorder
$\rho$ and an equilibrium configuration $\{\sigma_i\}$ at inverse
temperature $K_\rho$.  Then the disorder average of any quantity
$A(\{J_{ijk}\},\{S_i\},\beta)$ is equal to the disorder average of $A$
when $J_{ijk} \to J_{ijk} \sigma_i \sigma_j \sigma_k$ and $S_i \to
S_i\sigma_i$. To prove it consider another gauge transformation \bea
S_i &=& \tilde S_i \tau_i \, , \label{st}\\
\sigma_i &=& \tilde \sigma_i \tau_i \, , \\
J_{ijk} &=& \tilde J_{ijk} \tau_i \tau_j \tau_k \label{Jttt} \, .
\eea
Note that the Hamiltonian, but also $J_{ijk} \sigma_i \sigma_j \sigma_k$ and $S_i\sigma_i$ are invariant under (\ref{st}-\ref{Jttt}). 
Hence applying (\ref{st}-\ref{Jttt}) and averaging over all choices of $\tau_i\in \{\pm 1\}$ we get
\bea &&
\left[\<A(\{J_{ijk} \sigma_i \sigma_j \sigma_k \},\{S_i\sigma_i\},\beta)  \>_{K_{\rho}} \right]_{\rho} \nonumber
= \sum_{\{J_{ijk}\}} \\ \nonumber 
&&
 \frac{ \sum_{\{\tau_i\}}  \e^{K_{\rho} \sum_{ijk} J_{ijk}
    \tau_i \tau_j \tau_k }}{2^N 2^M (\cosh{K_{\rho}})^M}
\<A(\{J_{ijk} \sigma_i \sigma_j \sigma_k \},\{S_i\sigma_i\},\beta)  \>_{K_{\rho}} 
\nonumber
\\ 
&& = \frac {1}  {2^N(\cosh K_{\rho})^{M}} \sum_{\{J_{ijk}\}} \frac {Z(\{J_{ijk}\},K_{\rho})} {2^M}  \nonumber\\
&& \<A(\{J_{ijk} \sigma_i \sigma_j \sigma_k \},\{S_i\sigma_i\},\beta)  \>_{K_{\rho}}\, .
\label{NL-eq:4}
\eea
Using (\ref{blabla}), this proves relation~(\ref{theorem}). 

In particular, on the Nishimori line, i.e. when $K_{\rho}=\beta$ and $\rho_{\rm  NL} = 1/(1+e^{-2\beta})$, one has the following identity
\bea &&[A(\{J_{ijk}\},\{S_i\},\beta)]_{\rho_{\rm NL}} = \nonumber \\ &&\left[\<A(\{J_{ijk} \sigma_i \sigma_j \sigma_k \},\{S_i\sigma_i\},\beta)  \>_{\beta} \right]_{\rho_{\rm NL}}\, .
\label{THEOREM-BIG}
\eea
Thus on the Nishimori line all quantities behave the same if all spins
and all interactions are multiplied by factors $\sigma_i$
corresponding to an equilibrium configuration $\{\sigma_i\}$. Note
that the above disorder-averaged identities hold for any system, in
any dimension, with any number of spins. Since these disordered
systems are self-averaging in the thermodynamic limit (both in the
mean field \cite{Self-1,Self-2}, and in finite-dimension
\cite{Self-3}) identity~(\ref{THEOREM-BIG}) is also valid in the
thermodynamic limit on the Nishimori line even for {\it one given
  realization of the disorder}.

\subsection{Identities on the Nishimori line}
Let us give several specific examples of the generic identities one
can obtain on the Nishimori line. Consider the equilibrium energy per spin
\be 
e(\beta) = -\frac{1}{N}  \sum_{\{S_i\}} \left( \sum_{ijk}J_{ijk} S_i S_j S_k \right) \frac {e^{-\beta{\cal{H}}(\{\vec S\})}} Z \, .
\ee
Using (\ref{blabla}) and computing explicitly the annealed averages one obtains
\be
[e(\beta)]_{\rho_{\rm NL}} = - \frac MN \tanh{(\beta)}\, . \label{NL_energy}
\ee
This was first shown by Nishimori in \cite{Nishimori-Original}. Note
that this is nothing but the average energy of a configuration with
all spins up. The fact that the energy is an analytic function of
temperature implies that if a phase transition is present along the
Nishimori line then it is purely entropic, i.e. all non-analyticities
in the free energy have to stem from non-analyticities in the entropy
itself.

Let us now consider the average magnetization. One obtains from (\ref{THEOREM-BIG}) that
\bea m\! =\! \[[ \frac 1N \sum_{i} \langle S_i
\rangle \]]_{\rho_{\rm NL}} \!\!\!\!\! =\!\! \[[ \frac 1N \sum_{i}
\langle \langle\sigma_i S_i\rangle \rangle \]]_{\rho_{\rm NL}} =q_{EA}\, ,
\eea
where the double thermal average $\langle \langle\sigma_i S_i\rangle
\rangle$ is both over spins $\{\sigma_i\}$ and $\{S_i\}$, hence the
r.h.s is the equilibrium overlap $q_{EA}$ (or Edwards-Anderson
parameter \cite{Hertz}). The equilibrium magnetization on the
Nishimori line is equal to the average equilibrium overlap (which is
the standard order parameter in spin glasses) on the Nishimori line, a
fact again first shown by Nishimori \cite{Nishimori}.

Another useful identity is obtained when considering the so-called
Franz-Parisi (FP) potential \cite{FranzParisi}. It is a useful tool to
understand the properties of glassy states and it is defined as the
free energy of a system at temperature $T$ that has a fixed overlap
$q$ with a reference configuration $\{\sigma_i\}$ that is one of the
equilibrium configurations at temperature $T'$
\be
f_{\rm FP} (\beta,\beta',q) = \[[ \frac{ \sum_{\{\sigma_i\}}  e^{-\beta' {\cal H}(\{\sigma_i\})} f(\beta,\{\sigma_i\},q)}{\sum_{\{\sigma_i\}}   e^{-\beta' {\cal H}(\{\sigma_i\})}}\]]_{\rho}\, .
\ee
Under mapping (\ref{theorem}), we see that the
potential at temperature $T$ with respect to equilibrium at
temperature $T'$ is simply the free energy of a system with
temperature $T$ at fixed magnetization with disorder
$\rho=\rho_{\rm NL}(\beta')$. Indeed let $A$ be the free energy at fixed magnetization 
\be e^{-\beta N f(\beta,m)}=\sum_{\{S_i\}} \delta\(( \frac 1 N \sum_i
S_i - m\))\, e^{-\beta {\cal H}(\{S_i\})}\, , \ee by the mapping
(\ref{theorem}) we have \bea && [ f(\beta,m)]_{\rho_{\rm NL}(\beta')}
= \Bigg[ \frac{ \sum_{\{\sigma_i\}} e^{-\beta'
    {\cal{H}}(\{\sigma_i\})}
}{ Z(\beta') } \nonumber \\ && \log{  \left\{\sum_{\{S_i\}} \, \delta(\frac 1 N\sum_i \sigma_i S_i - m) \,  e^{-\beta {\cal{H}}(\{S_i\})}  \right\} }  \Bigg]_{\rho_{\rm NL}(\beta')} \nonumber  \\
&&= [ f_{\rm FP}(\beta,\beta',m)]_{\rho_{\rm NL}(\beta')} \, .  \eea
In particular, if $\beta=\beta'$ the Franz-Parisi potential is nothing
but the equilibrium free energy at fixed magnetization. This shows
that the physics relative to the fully magnetized configuration is
identical to the equilibrium physics: this is the particularity of the
Nishimori line and a crucial point for this paper.
 
One can apply the gauge transformation (\ref{gauge_s}-\ref{gauge_j}),
and the identity (\ref{theorem}), to dynamical quantities as well, as
was first realized in \cite{Ozeki1,Ozeki2}. For instance, Glauber,
heat-bath or any dynamical evolution that satisfies the balance
condition are gauge invariant (this is simply because the {\it
  Hamiltonian itself} is invariant, so that the dynamics is not
affected by the gauge transformation). Consider for the sake of the
discussion the heat-bath dynamics initialized in the fully
ferromagnetic configuration, $S_i(0)=1$, for a system on the Nishimori
line $\beta=K_\rho$. The probability that the magnetization evolving
under that dynamics has a certain value reads
\bea &&[P(m(t)|m(0)=1)]_{\rho_{\rm NL}} = \nonumber \\ && \Big[
\Big\langle P\big(\sum_i S_i(t) \sigma_i/N|\sum_i S_i(0) \sigma_i =N \big)
\Big\rangle_{\beta} \Big]_{\rho_{\rm NL}} \, .
\label{dyn-th}
\eea 
Hence, for the $p$-spin model on the Nishimori line, given a
Hamiltonian based dynamics, the decorrelation from an equilibrium
configuration follows the same functional form as the decorrelation
from the fully ordered configuration, a result first shown by
\cite{Ozeki1,Ozeki2}. Eq.~(\ref{dyn-th}) implies that melting dynamics
is equivalent to equilibrium dynamics: on the Nishimori line the
magnetization decay starting from the fully ordered configuration is
equal to the dynamical correlation function
 \bea
  m_{\rho_{\rm NL}}(t) &=&  \frac 1N \sum_i S_i(t)  \nonumber \\ 
= C_{\rho_{\rm NL}}(t) &=& \frac 1N \sum_i S_i(t_1) S_i(t_1+t) \, , \label{corr} \eea
if $S_i(t_1)=\sigma_i$ is an equilibrium configuration. One can push
this idea one step further and obtain interesting identities for the evolution
of the dynamic ferromagnetic susceptibility $\chi_F$. Let us denote $\{{.\}}_{dyn}$
the average over many realizations of the dynamics, then $\chi_F$ reads
 \be \chi_F(t)= \frac{\beta}{N} \sum_{ij}
\Big[ \{{ S_i(t)S_j(t) \}}_{dyn} -\{{ S_i(t) \}}_{dyn} \{{S_j(t) \}}_{dyn} \Big]_{\rho_{\rm NL}}\label{susc_F}
\ee Using the same transformation, we see  that $\chi_F(t)$
on the Nishimori line equals the equilibrium $4$-points
susceptibility $\chi_4(t)$ defined as
 \bea &\chi_4(t)&
= \frac{\beta}{N} \sum_{ij} \Big[ \{{ S_i(t_1+t)S_i(t_1)S_j(t_1+t)S_j(t_1)
\}}_{dyn} \nonumber \\ &-& \{{ S_i(t_1+t)S_i(t_1) \}}_{dyn}\{{
S_j(t_1+t)S_j(t_1) \}}_{dyn} \Big]_{\rho_{\rm NL}}  \label{susc_4}
\eea 
with $S_i(t_1)=\sigma_i$ being again an equilibrium configuration.  The
melting process on the Nishimori line thus satisfies\footnote{As noted
  by \cite{Silvioetal} there are other possible definitions of the
  dynamical susceptibilities, depending on the order of the
  averages, that lead to slightly different equalities on the
  Nishimori line. For instance, one could use: $ \chi_F(t)=\Big[ \{{
    m(t)^2 \}} \Big]-\Big[\{{ m(t) \}}\Big]^2=\Big[ \{{ C(t)^2
    \}}\Big] -\Big[\{{ C(t) \}} \Big]^2=\chi_4(t).$}
\bea
&&\chi^{\rho_{\rm NL}}_F(t) = \beta \Big[ \{{ m(t)^2 \}}_{dyn} -\{{
  m(t) \}}_{dyn}^2 \Big]_{\rho_{\rm NL}} \nonumber \\
&=&\chi^{\rho_{\rm NL}}_4(t) = \beta \Big[ \{{ C(t)^2 \}}_{dyn} -\{{
  C(t) \}}_{dyn}^2 \Big]_{\rho_{\rm NL}} \nonumber 
\label{susc}
\eea
The bottom line here is that the physics relative to the fully
magnetized state is at the same time the equilibrium physics so that
the melting dynamics from the fully ordered state is equivalent to the
equilibrium dynamics. 

\subsection{First order transitions on the Nishimori line}
In a companion article \cite{US-PART-I} we have discussed analogies and
differences between the equilibrium dynamics in super-cooled liquids
and the melting dynamics in a system with a general first order phase
transition.

To remind the main findings, we consider the melting dynamics in a
spin system with a first order phase transition. We initialize the
system in the fully ordered state (that is the completely magnetized
one) and suddenly change the temperature to put the system in the
paramagnetic phase. The system will melt into the less ordered
phase. This was discussed in detail in \cite{US-PART-I} (we also refer
the reader to the classical articles \cite{ReviewBinder,Binder73}),
and the phenomenology of this melting process is strikingly similar to
the one of the dynamics of super-cooled liquids.

In particular the melting time diverges super-exponentially as the
first order phase transition is approached, as can be understood by
standard nucleation arguments. In glasses the equilibration time also
grows super-exponentially, and according to some diverges at an {\it
  ideal} glass transition temperature. In a system with a first order
transition we also observed diverging dynamical and static correlation
lengths. In particular, the dynamical length scale
\cite{Silvio,DynBB,DynHS} is associated to heterogeneities in the
dynamics. It uses a four-point density correlator in both time and
space, and it led to the notion of the so-called dynamical
susceptibility (usually refereed to as $\chi_4$). The $\chi_4$
susceptibility was observed to grow also in glass formers
\cite{DynBB,DynHS}. There is also a {\it static} (thermodynamic)
growing length scale associated to a first order phase transition, the
so-called point-to-set correlation which is the correlation of a
sub-system with its frozen boundaries. This correlation length was
observed  to grow also in glass formers \cite{BiroliBouchaud,DynamicBethe,PointToSet}.

Despite the profund analogy \cite{US-PART-I} there are crucial
differences between glassy dynamics and the melting one through a
standard first order transition. Consider thermodynamic properties:
there is no latent heat associated to the glass transition, as the
energy at a glass transition is continuous, while in general there is
a latent heat in first order phase transitions. This difference can be
overcome by considering a first order transition driven purely by
entropy where there is no latent heat, as for instance in hard spheres
\cite{HardSphere}, liquids crystal and other systems \cite{Frenkel}. A
more serious difference between glassy dynamics and standard melting
appears in the dynamical behavior. Usually the melting process happens
once for all and when the system has melted, it stays in the liquid
phase, usually a very different one from the ordered phase. The
equilibrium glassy dynamics, on the other hand, is a stationary
process which is time-translationally invariant.

Considering the results of the previous sections, one sees that both
these differences vanish on the Nishimori line: the melting dynamics
is equivalent to the stationary equilibrium dynamics. Moreover, there
is no latent heat as the energy is given by (\ref{NL_energy}).  More
importantly, if there is a first order phase transition on the
Nishimori line then there {\it are} divergent length and time scales,
and the mapping between melting and equilibrium dynamics tells us that
there are genuine divergences in the equilibrium dynamics as well! To
conclude: if there is a first order phase transition on the Nishimori
line then we found an exact and simple-to-study realization of the
ideal glass transition. This is the main thesis of this paper and we
shall now pursue it, first in a mean-field system, and then in a
finite dimensional one.

\section{Glassy dynamics as a melting process in mean field systems}
\label{Sec:2}
We shall first study the $p$-spin model on the Nishimori line in the
mean-field setting. As we shall see, this will be equivalent to the
usual spin-glass mean field theory, so it is useful to first review
the properties of mean-field spin glasses. We will consider a random
lattice where every spin is involved in exactly $c=O(1)$
interactions. Such a diluted $p$-spin model is called the XOR-SAT
problem in the literature and can be solved using the replica or the
cavity method (see \cite{XORSAT,Following,FollowingLong}).  We chose
to work with the diluted $p$-spin model instead of the more common
fully connected one \cite{KT} because it can be simulated in a
time linear (versus quadratic) in the size of the system,  and because
a distance between spins is naturally defined. 

The Ising $p$-spin glass \cite{REM,P-SPIN,XORSAT}, that is Hamiltonian
(\ref{PSPIN-H}) when $\rho=1/2$, provides a mean-field theory for the
structural glass transition \cite{KT}. The thermodynamic behavior
of the $p$-spin glass undergoes the following changes as the
temperature $T$ is decreased (see Fig.~\ref{FigPHASE}): At high
temperature, the system is in a paramagnetic (or liquid) phase. Below
the so-called {\it dynamical} glass temperature, that we denote here
$T_{\rm MCT}$, the paramagnetic state shatters into exponentially many
Gibbs states: the energy landscape is therefore divided into
exponentially many dynamically attractive regions, all well separated
by extensive free-energetic barriers. This leads to a breaking of
ergodicity on any non-exponential time-scale and to the power-law
divergence of the equilibration time at $T_{\rm MCT}$
\cite{Kurchan,DYNAMIC,AndreaSaddle,DynamicBethe}. This is the
equivalent of the mode-coupling transition in structural
glasses. Note, however, that this dynamical transition is not a phase
transition in the usual sense: there is no non-analyticity in the free
energy at the transition. It is only a topological transition in the
configurational space that affects the dynamics of the system (thus
the name {\it dynamical transition}).

As the temperature is further lowered, the number of states (relevant for
the Boltzmann measure) decays. A~second {\it static} Kauzmann
transition is then reached at $T_K$ when the number of relevant states
becomes sub-exponential (and in fact finite) and the structural
entropy (or complexity) vanishes. The RFOT departs from
this mean-field picture and argues that in finite dimensional systems
$T_{\rm MCT}$ is only a crossover so that the relaxation time diverges
at $T_K$. But in this section we will restrict ourselves to the
mean-field case.

Another property of the $p$-spin model with $\rho=1/2$, that will turn
out to be very useful in what follows, is that the {\it annealed}
averages are equal to the {\it quenched} ones above the Kauzmann
temperature.

\subsection{Annealed and quenched averages in the mean field $p$-spin model} 
In the quenched average the disorder realization (i.e. the lattice and
signs of interactions) is fixed, and the thermodynamic average over
configurations is taken. Only after that the disorder average is taken, and this is the physically correct way to proceed in most disordered
systems. In the annealed case (which is in general only approximate)
the average over disorder is taken always {\it at the same time} as
the average over configurations.

In the $p$-spin model above the Kauzmann transition, $T_K$, the
quenched free energy is asymptotically equal to the annealed one
$-\beta f=\left[\log{\((Z\))}/N \right]_{1/2}=
\log{\left[Z\right]_{1/2}}/N + o(1)$, see \cite{P-SPIN,XORSAT}.
Assuming at least exponentially rare large deviations for thermodynamic
quantities, we can see that choosing first a disorder realisation and
then a random configuration of a given energy is the same as taking a
random configuration and choosing the disorder at random such that the
configuration has the same energy. This is because in the second case
instances of disorder are chosen proportionally to the value of the
partition function at a corresponding temperature but since
$\left[\log{\((Z\))} \right]_{1/2}= \log{\left[Z\right]_{1/2}}+o(N)$
this second way chooses typical instances as well\footnote{Notice that
  equality of the quenched and annealed average holds for a larger
  class of problems than just the $p$-spin model, see for instance the
  models where a {\it quiet planting} is possible
  \cite{AchlioptasCoja-Oghlan08,KrzakalaZdeborova09,ZdeborovaKrzakala09}.}.

In what follows in the mean field $p$-spin model we can thus freely exchange the quenched average of
thermodynamic quantities for annealed one or vice
versa, i.e.
\bea \left[ \langle A \rangle_\beta \right]_{\frac{1}{2}}\equiv \left[
  \frac { \sum_{\{{S\}}} A e^{-\beta {\cal H}}}{Z}
  \right]_{\frac{1}{2}} = \frac {\left[ \sum_{\{{S\}}} A e^{-\beta
      {\cal H}}\right]_{\frac{1}{2}}} {\left[ Z \right
  ]_{\frac{1}{2}}} \, .
\label{e1}
\eea
We stress that the above holds only for disorder averages of
thermodynamic quantities (i.e. when the quantity or its density is
bounded by an $N$-independent constant), for instance it does not have
to hold that $\left[Z\right]_{1/2}^2 = \left[Z^2\right]_{1/2}$. We now
denote the number of interaction by $M$ and compute explicitly the
free energy. First, we obtain
\be
[Z]_{\frac{1}{2}}= \frac 1{2^M} \sum_{\{J_{ijk}\}} \sum_{\{{S_i\}}}
e^{-\beta {\cal{H}}}= 2^N \cosh^M{(\beta)} \, . \ee
The free energy density for $T \ge T_K$ is thus simply
\be
- \beta f = \log{2} + \frac MN \log{(\cosh{\beta})} ,
\ee
and the energy is given by
\be
e = - \frac{{\rm d} (\beta f)}{{\rm d}\beta} = \frac MN \tanh{(\beta)}\, ,
\ee
as we could also obtain from eq.~(\ref{e1}). These results are correct
for all $T\ge T_K$. Just to be concrete, for the $3$-spin model where
every spin is involved in $c=5$ interactions, one finds $T_K = 0.849$,
while the system is in the many valleys glassy phase for all $T_K\le
T<T_{\rm MCT}=0.936$.

\subsection{The Nishimori line in a mean field $p$-spin model}
\label{mf-nish}
Let us now consider the general identity (\ref{blabla}), when the quenched
and annealed averages at inverse temperature $K_\rho$ are equal we
arrive to
\be 
[A(\{{J_{ijk}\}},\{S_i\},\beta)]_{\rho}  =
\left[\langle A(\{J_{ijk} \sigma_i \sigma_j \sigma_k \},\{S_i\sigma_i\},\beta) \rangle_{K_\rho}
\right]_{\frac{1}{2}}\, ,
\label{theorem-meanfield}
\ee
where $\langle\cdot\rangle_{K_\rho}$ is the thermal average at inverse
temperature $K_\rho$. This means that physics of the mean-field spin glass at temperatures $\beta=K_\rho$ is uniquely
mapped on physics of the same model but with a ferromagnetic
bias $\rho_{\rm NL}$ corresponding to the Nishimori line. This holds above the Kauzmann temperature, $T\ge T_K$, and in any
model where the annealed average equals the quenched one (that
does not include any finite-dimensional model we are aware of).

We give several more specific examples of the above
statement. According to the previous section the melting process on
the Nishimori line is equivalent to the equilibrium dynamics on the
Nishimori line. In the mean field $p$-spin moreover it is equivalent
also to the equilibrium dynamics in the spin glass problem,
$\rho=1/2$, as long as $T\ge T_K$. One has
\begin{align}
&m_{\rho_{\rm NL}}(t)=C_{\rho_{\rm NL}}(t)=C_{\rm SG}(t)\, , \label{m_sg}\\
&\chi_F^{\rho_{\rm NL}}(t)=\chi_4^{\rho_{\rm NL}}(t)=\chi_4^{\rm SG}(t)\, , \label{chi_sg}\\
&f^{\rho_{\rm NL}(\beta')}(\beta,m)=f_{\rm FP}^{\rho_{\rm NL}(\beta')}(\beta,\beta',m)= f_{\rm FP}^{\rm SG}(\beta,\beta',m)\, .
\end{align}
The above mapping allows us to understand in a simple manner the
glassy behavior of the $p$-spin model\footnote{It also simplifies
  many analytical computations. The FP potential computed in the
  spherical $p$-spin \cite{FP-spherical} can for instance be obtained by considering the free energy with a ferromagnetic bias
  \cite{FP-Sherrington}. In the case of Ising spins the solution can be obtained readily by looking to the free energy on the Nishimori line
  \cite{NishimoriWong}. This was discussed by the present authors
  in \cite{Following,FollowingLong}.}. Let us consider an equilibrium configuration
at temperature $T>T_K$, the time needed to decorrelate diverges at the
dynamical, or mode-coupling transition $T_{\rm MCT}$.  The correlation
function (\ref{corr}) develops a plateau whose length diverges at
$T_{\rm MCT}$. Using the above mapping, these features translate in
the system on the Nishimori line. On the Nishimori line there is a a
first order phase transition at $T_F=T_K$ from a paramagnetic to a
ferromagnetic phase with the following properties.  Initializing the
dynamics in the fully ordered configuration, the magnetization decays
to zero only for temperatures larger than the spinodal one $T_s=T_{\rm
  MCT}$ (since for $T<T_s$ the system will be trapped in the
ferromagnetic state). The time needed to relax to zero magnetization
diverges at $T_s$ as a power law (since we deal with a mean-field
system). The magnetization $m(t)$ develops a plateau whose length
diverges at $T_s$.

\subsection{Simulating glassy dynamics as a melting process}
\label{simu-super}
We shall now simulate the melting process on the Nishimori line. A
useful application of the idea discussed above is that it provides
efficient numerical simulations of the equilibrium dynamics in the
spin glass. Instead of equilibrating Monte-Carlo simulation in the
spin glass model we can simply initialize in the fully magnetized
configuration on the Nishimori line.

\begin{figure}[t]
\hspace{-0.6cm}
\includegraphics[width=9cm]{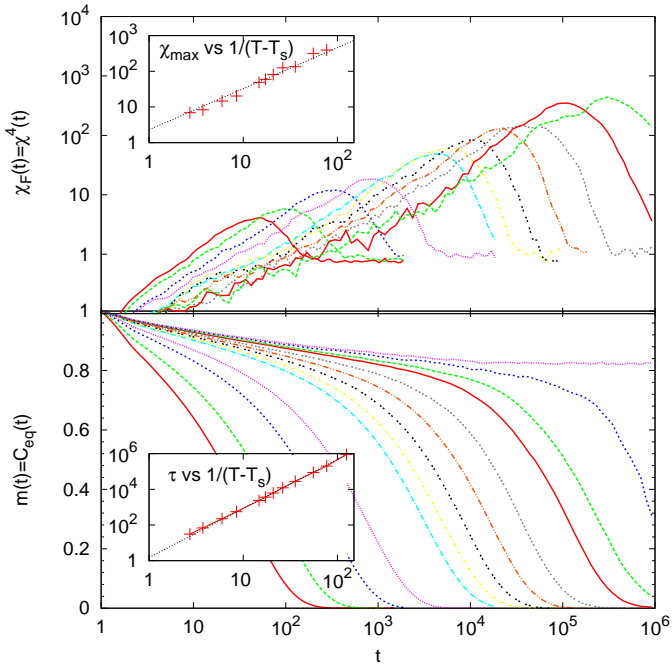}
\caption{Dynamic ferromagnetic susceptibility and decay of the
  magnetization $m(t)$ in the melting on the Nishimori line for the
  $3$-spin model on the Bethe lattice with $c=5$ and $N=90000$ spins,
  averaged over $500$ realizations of the dynamics except for the
  three lower temperature where we used only $50$ runs.  From left to
  right: $T=1.3$, $1.2$, $1.1$, $1.05$, $1.0$, $0.99$, $0.98$, $0.97$,
  $0.96$, $0.955$ and $0.95$. Bottom inset: the melting time [defined
  by $m(\tau)=1/e$] that diverges as a power law (\ref{tau_p}).  Top
  inset: The maximum of the ferromagnetic susceptibility $\chi_F$
  diverges as (\ref{chi_max_p}).  The plotted quantities are identical
  to the dynamical susceptibility $\chi_4(t)$, eq.~(\ref{chi_sg}), and
  the equilibrium autocorrelation $C(t)$, eq.~(\ref{m_sg}), in the
  spin glass phase. \label{dyn_mf} }
\end{figure}

We have performed Monte-Carlo simulations of the melting process of a
large system on the Nishimori line. For the $3$-spin model with $c=5$
for temperature $\beta$, we have prepared a random realization of the
system with $N=90000$ spins and a proportion of negative couplings
given by the Nishimori condition eq.~(\ref{nishimori}), and
initialized the simulation with all spins $S_i=1$. We then use
Monte-Carlo dynamics with the Metropolis update rule. The results are
shown in Fig.~\ref{dyn_mf}. The exact mapping ensures that the
magnetization evolves the same way as the correlation function in the
spin glass, and that the dynamic susceptibility evolves as the 4-point
correlation function in the spin glass. The relaxation time diverges
as
\be \tau \propto \left(1-\frac{T_{\rm MCT}}{T} \right)^{-2.8(2)}\,
, \label{tau_p} \ee
where the critical exponent was obtained by fitting the curve in the
inset of Fig.~\ref{dyn_mf} and agrees well with the values obtained in
\cite{DynamicBethe}. The maximum of the dynamical susceptibility
diverges as \be \chi_{\rm max} \propto \left(1-\frac{T_{\rm MCT}}{T}
\right)^{-1.15(15)}\, , \label{chi_max_p} \ee where the critical exponent
was obtained by fitting the curve in the inset of
Fig.~\ref{dyn_mf}. This agrees quite well with the predictions of the
mode-coupling theory \cite{MCT,MCT2,BiroliBouchaudMCT,DynBB} where the
exponent is equal to $-1$.

Let us stress that this simulation would take a much longer time
without the above trick that gives us an equilibrium configuration
{\it for free}. Moreover, the method extends to temperatures $T_K \le T
\le T_{\rm MCT}$ where Monte-Carlo equilibration takes an exponential
time and is hence infeasible for such large system sizes\footnote{Note
  that the authors of \cite{DynamicBethe} used a similar trick by
  simulating the annealed model.}.

\begin{figure}[t]
\vspace{-0.1cm}
\hspace{-0.6cm}
\includegraphics[width=9cm]{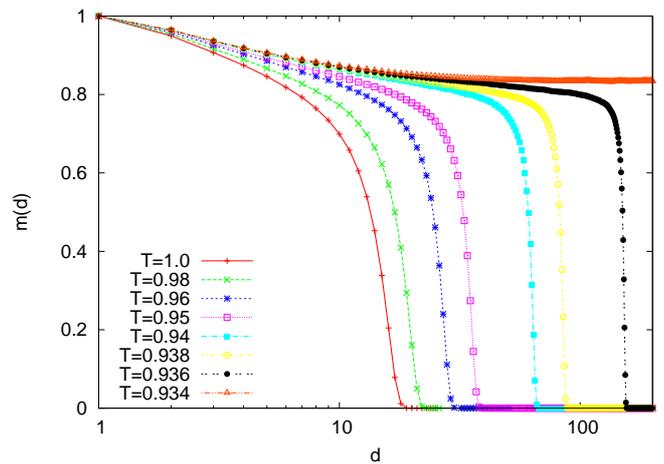}
\caption{Magnetization at distance $d$ from a ferromagnetic boundary
  conditions in $3$-spin model on a Bethe lattice of coordination
  $c=5$ on the Nishimori line. This is the point-to-set correlation in
  the spin glass case. The correlation length diverges with exponent
  $1/2$ at $T_{\rm MCT}=0.936$. \label{PTS-MF}}
\end{figure}

The above mapping also allows to understand the point-to-set
correlations \cite{DynamicBethe,PointToSet} in a simpler
way. Point-to-set correlations are defined as follows: freeze the
system in an equilibrium configuration. Then consider a large compact
droplet in the system and un-freeze it to observe the influence of the
frozen boundaries. For each temperature, one defines the equilibrium
point-to-set volume as the largest size of the droplet such that the
boundaries are correlated with the center of the droplet. Using the
above mapping, this notion translates to a much simpler, and more easy
to simulate, one: we just
have to consider a system with fully ferromagnetic boundary conditions
and ask if the magnetization stays positive in a center of a droplet
of a certain size: This is the very same question we addressed in the
purely ferromagnetic system in \cite{US-PART-I}.

Point-to-set correlations in the mean-field spin glass case can be
derived analytically and obey a simple solvable recursion (see
\cite{DynamicBethe}). Our mapping shows (as can also be seen
explicitly in the equations) that this recursion is {\it identical} to
the one describing the way the magnetization decays at distance $d$
from ferromagnetic boundary conditions on the Nishimori line. We plot
the solution to these equations in Fig.~\ref{PTS-MF}. Actually since
the equations in question are very similar to what happens in pure
ferromagnets, it is not surprising that the decorrelation length grows
when $T_{\rm MCT}$ is approached with the exponent $1/2$
that was first obtained, for this model, in \cite{DynamicBethe}.

\section{Glassy dynamics as a melting process in a finite dimensional
  model}
\label{Sec:3}
We now discuss how is this mean field picture modified in finite
dimensional systems. Since we are dealing with a glass transition, we
should in principle use the mosaic theory, which is the counter-part
of nucleation theory for glasses
\cite{AdamGibbs,KW1,KW2,KTW,BiroliBouchaud}. The fundamental point in this
work is that due to the mapping to melting above a first order phase
transition (from a ferromagnetic phase to a paramagnetic one) it is
sufficient to consider the standard nucleation theory. This is a great
conceptual simplification with respect to the many studied lattice models of
glass formers \cite{Plaquette_L,Plaquette_B,STAR_model}.

\subsection{Nucleation theory for the melting of a disordered magnet}
\label{sec:nucl}
Let us very briefly recall the basics of nucleation theory, and
discuss the effect of disorder for an entropy-driven first order phase
transition. Nucleation stems from a competition between the bulk {\it
  entropy} difference and the surface tension between two phases.
When a system is in a metastable phase the total free-energy cost
$\Delta F(\ell)$ of a compact droplet of size $\ell$ is given by a
combination of the bulk entropy $-\ell^d T \delta s$ and the surface
tension $\ell^{\theta} \Gamma$, where $\delta s$ is the entropic
difference between the two phases, $\theta\le d-1$ is an exponent
characterizing the free energy cost of the surface, and $\Gamma$ is
the surface tension (we omit constant pre-factors for convenience).
In a non-disordered systems we have $\theta=d-1$, but when disorder is
present the shape of the droplet can adapt in order to take advantage
of the local configuration of the disorder, so for the {\it best}
droplet of each size we have $\theta \le d-1$. The two contributions
balance when the droplet size is
\be \ell_{\rm c} \propto {\left( \frac{\Gamma}{T \delta s}
  \right)}^{\frac 1 {d-\theta}} \propto (T-T_c)^{-\nu}
\label{3D-nucl}
\ee
with $\nu=1/({d-\theta})$, where we assumed that the difference
$\delta s$ is linear when approaching the first order phase
transition. For sizes larger than $\ell_c$ flipping a droplet is
advantageous for the system.  In order to flip, the thermal
fluctuations must cross the free energetic barrier. In a
non-disordered system one would expect that the barriers grow as
$\ell_c^{\theta}$. However, this is only the free energy of the best
droplet along the way, but the path that the dynamics is following is
not bound to pass through this {\it best} droplet; it might actually
have to go through larger barriers, and it is thus safer to assume
instead that the barriers grow as $\ell^{\psi}$ with $\psi \ge
\theta$. Using the Arrhenius formula we thus obtain the nucleation
time
\be \tau \propto \exp \left[\frac{A}{T (T\delta s)^{\gamma}} \right] =
\exp\left[\frac{\tilde A}{(T-T_c)^{\gamma}} \right]
\label{3D-TIME}
\ee
with $\gamma=\psi/{(d-\theta)}$.

So far we have, however, neglected an important ingredient: disorder
induced fluctuations. We shall now present a scaling analysis in order
to estimates the effect of the disorder on a first-order transition.
Consider a droplet of size $\ell$, we expect that it can gain a free
energy of order $\Delta F_{\rm dis} = \delta J \ell^{d/2}$ just from
fluctuations in the free energy due to the disorder variance $\delta
J$. The point is that this might be enough to make the droplet flip if
the disorder induced term is larger than the surface term $\Gamma
\ell^{\theta}$. Indeed, if $\theta <d/2$ there is a {\it finite}
length $\xi_{\rm dis}\propto \(({{\Gamma}/{\delta
    J}}\))^{1/(d/2-\theta)}$ beyond which the droplet flips
independently of temperature. Hence in such a case the critical
droplet size (that is, the point-to-set size) does not diverge at the
transition point, and consequently the transition {\it cannot} be of
first order. If we want a genuine first order phase transition, we
thus must have $\theta \ge d/2$, or equivalently
\be
\nu=1/{(d-\theta)} \ge 2/d.
\label{condition}
\ee 
For instance, the introduction of small amount of disorder in the pure
ferromagnetic model, where $\theta=d-1$, cannot lead to a first order
phase transition for $d<2$. The situation is the same in the marginal
case $d=2$, as proven rigorously by Aizenmann and Wehr~\cite{Self-3}.
According to this analysis, we thus need at least a three-dimensional
model if we want to have a hope for a genuine first order phase
transition in the presence of disorder.

Finally, following the discussion from a companion paper
\cite{US-PART-I}, let us emphasize that the relevant mechanism for
melting is nucleation {\it and} growth. The nucleation time $\tau$
(\ref{3D-TIME}) only gives the average time needed to nucleate a
critical droplet {\it at a given position} in the system.  If the
system is much larger than the critical nucleus size the melting
process is {\it not} done via the growth of a single droplet but
instead results from the nucleation of {\it many} droplets that will
simultaneously grow and invade the system. This growth time has to be
added to the nucleation time to obtain the relaxation time needed to
exit from the metastable state in a very large system. The probability
to nucleate a critical droplet by a unit of volume and time is
proportional to $e^{-\beta \Delta F(l_{c})} \propto 1/\tau$.  Hence in
a very large system there is roughly one droplet per volume $\tau$
after a unit of time. Without disorder, the volume of large droplets
is expected to grow algebraically with time and consequently in the
systems without disorder the total relaxation time is of the same
order as the nucleation time $\tau$, and the typical size of grown
droplets when they percolate is also scaling  as $\tau$. This can
be observed by looking to the dynamical ferromagnetic susceptibility,
defined in eq.~(\ref{susc_F}), as we have discussed in detail in
\cite{US-PART-I} for the Potts model. Clearly, the moment when such
droplets percolate denotes the moment when the system is most
heterogeneous, where roughly half of it is in the liquid phase and the
other part still in the ordered one, so that the two-point dynamical
ferromagnetic susceptibility is maximal.

Disorder will modify the growth behavior. Instead of the algebraic
nucleus growth we expect instead an activated growth due to pining of
the interfaces, just like in the coarsening of disordered magnets
where the growth is logarithmic in time \cite{RFIM-th}. The interplay
between activated growth and the appearance of new droplets makes the
total analysis more involved than in the case without disorder, and is
a subject worthy of closer inspection in order to understand the
melting of a crystalline ordered phase in presence of disorder. If the
growth is slow, the melting will have to wait until many droplets have
nucleated everywhere. Again, this gives a time of order $\tau$ for the
melting process, but when droplets percolate, they are typically much
smaller than in the non-disordered case. We will see that the
dynamical magnetic susceptibility is again a suitable tool to quantify
this understanding.

\subsection{Nucleation vs mosaic on the Nishimori line}
We now focus on a first order phase transition {\it on the Nishimori
  line}. Since we are discussing an entropy driven transition in a
system with disorder, the analysis of the former paragraph applies. On
the Nishimori line, however, the melting process is equivalent to the
equilibrium dynamics. That means the system that is in a given
equilibrium configuration decorrelates from this configuration by the
nucleation mechanism to reach another equilibrium configuration which
again will be left by activated nucleation and so on. Such a
stationary melting process, where one transits from one equilibrium
configuration to another by nucleation is nothing else than the
phenomenology of the RFOT and the mosaic picture. Due to our mapping,
however, it simply appears as a consequence of standard nucleation
arguments.

For instance, eq.~(\ref{3D-TIME}) is nothing but the Adam-Gibbs
relation \cite{AdamGibbs,KW1,KW2,KTW,BiroliBouchaud}, while
eq.~(\ref{3D-nucl}) is the mosaic length. The entropy difference is
now associated with the number of metastable state so that $\delta
s=\Sigma$ (where $\Sigma$ is the configurational entropy).  According
to the condition (\ref{condition}), we need $\nu\ge 2/d$, a conclusion
was also reached in the seminal article on the RFOT \cite{KTW} where
in fact it was even argued that $\nu=2/d$, which combined with
$\psi=\theta$ gives $\gamma=1$, which gives the original Adam-Gibbs
relation. If indeed $\nu=2/d$ it suggests that if there is a
first-order transition on the Nishimori line, it is a marginally
stable one with respect to the smoothing by disorder. These questions,
and in particular whether indeed $\nu=2/d$ or not, deserve further
investigations.

An interesting (and hopefully clarifying) relation in our mapping is
given by eq.~(\ref{susc}). The four-point dynamical susceptibility
$\chi_4(t)$, that describes the heterogeneous glassy dynamics, is
simply the dynamical ferromagnetic susceptibility in the melting
process. $\chi_4(t)$ thus acquires a simpler interpretation, as
discussed in section \ref{sec:nucl}. The mapping on the Nishimori line
thus allows to discuss the dynamics of glass forming liquids as a much
simpler melting process. Let us illustrate this in a three-dimensional model.

\subsection{A finite dimensional $p$-spin model}
There are many ways how to introduce a $p$-spin model on a finite
dimensional lattice (see for instance \cite{P-SPIN-FINITE}). We will
use the following model on a three-dimensional cubic grid with each
spin being involved in $10$ different $5$-body interactions
\bea 
{\cal H}=& -&\sum_{i} J^a_{i} S_{i} S_{\rm up} S_{\rm left} S_{\rm right} S_{\rm back}
\nonumber \\
&-& \sum_i J^b_{i} S_{i} S_{\rm down} S_{\rm left} S_{\rm right}
S_{\rm front} \, ,
\label{3D}
\eea
where the indexes indicate the position of the spin on the grid with
respect to the spin $i$. In the mean field case ---that is on a Bethe
lattice with $p=5$ and $c=10$--- this model can be solved using the
cavity method and it has qualitatively the phase diagram from Fig.~\ref{FigPHASE}, with a first order phase transition at
$T_c(\rho=1)=2.4$ in pure case while on the Nishimori line,
$e^{2\beta}=\rho/(1-\rho)$, we have $T_c(\rho_{\rm NL})=0.96$. 

\subsection{Numerical results}
\begin{figure}[ht]
\vspace{-0.1cm}
\hspace{-0.6cm}
\includegraphics[width=9cm]{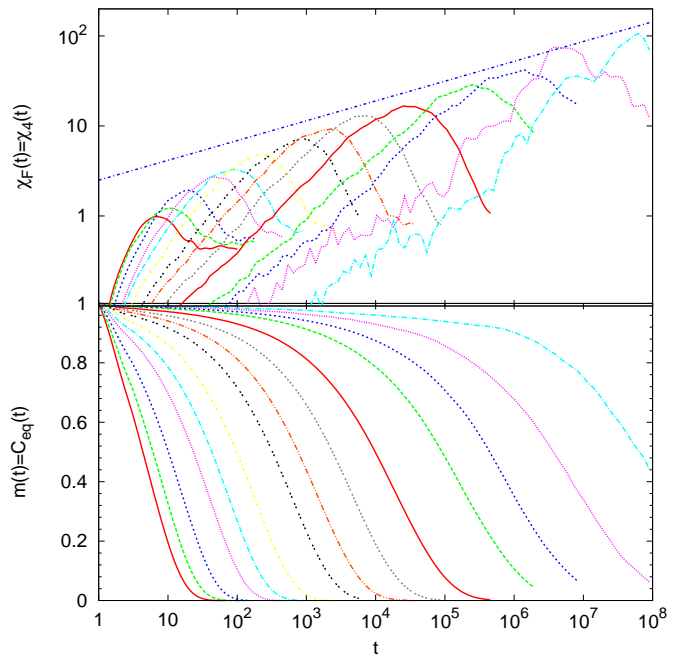}
\caption{The dynamics on the Nishimori line for the $3$-dimensional $5$-spin model. Top: Dynamical susceptibility or equivalently the 4-point correlation function, $\chi_4(t)=\chi_F(t)$, as a function of time. Bottom: The magnetization or equivalently the equilibrium autocorrelation function, $C(t)=m(t)$, as a function of time. We use $1000$ realizations of
the dynamics, except for the two lowest temperatures where we used only
$50$ realizations. From left to right: $T=2.2$, $2.0$, $1.8$, $1.6$,
$1.5$, $1.4$, $1.3$, $1.25$, $1.2$, $1.15$, $1.1$, $1.075$, $1.05$
and $T=1.025$. The peak of the dynamical susceptibility seems to increase
slower than a power law of the relaxation time $\tau$. We show  a power-law line with
exponent $0.22$, which is the best power-law estimate for the largest
time; however such a low value, and the general trend of the maxima,
suggest that asymptotically the growth is logarithmic in $\tau$ (and thus
algebraic in $T-T_c$) as expected for an actived coarsenning.
\label{3D-SIMU1}}
\end{figure}

\begin{figure}[ht]
\vspace{-0.1cm}
\hspace{-0.6cm}
\includegraphics[width=9cm]{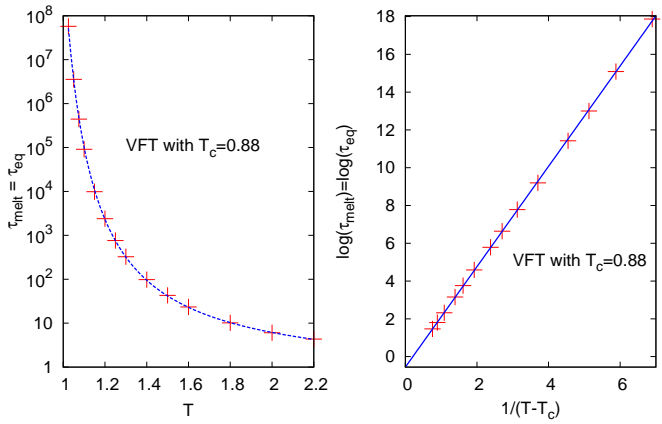}
\caption{Melting time or equivalently the relaxation time $\tau$, in
  the $3$-dimensional $5$-spin model as a function of the temperature
  $T$ (plotted in a log-linear way on the left, and logarithm of the
  time versus $1/(T-T_c)$ on the right). The dependence is very well
  fitted with a Vogel-Fulcher-Talman form $\tau \propto
  e^{\frac{A}{(T-T_c)}}$, with $T_c \approx 0.88$. \label{tau_3d} }
\end{figure}

\begin{figure}[th]
\vspace{-0.1cm}
\hspace{-0.6cm}
\includegraphics[width=9cm]{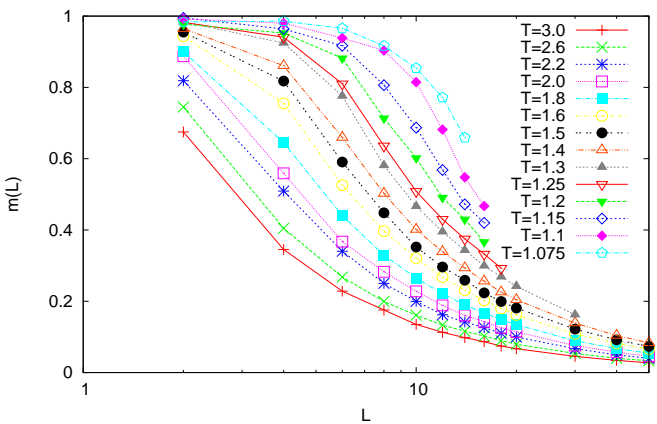}
\caption{Point-to-set correlation, or equivalently equilibrium
  magnetization $m(T,L)$ in a box of size $L^3$ with ordered boundary
  conditions, for different temperatures. A clear sign a growing
  length scale as the temperature is lowered is
  observed. \label{3D-PTS}}
\end{figure}

\begin{figure}[th]
\vspace{-0.1cm}
\hspace{-0.6cm}
\includegraphics[width=9cm]{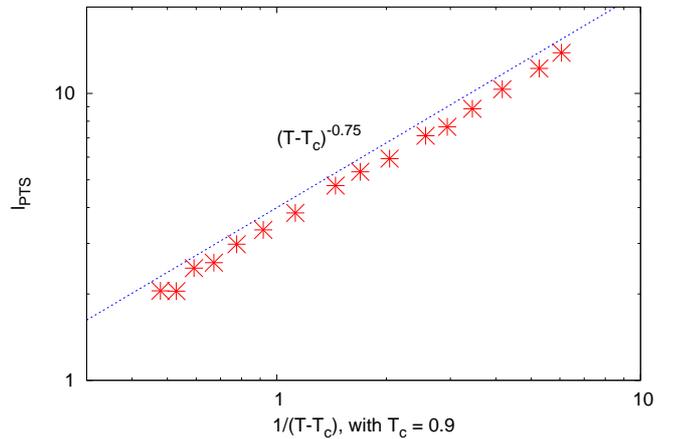}
\caption{Point-to-set correlation length, defined as the size of the
  box where the magnetization in Fig.~\ref{3D-PTS} goes below
  $2/3$. It is well fitted by $1/(T-T_c)^{\nu}$ with $\nu \approx 0.75$
  and $T_c=0.9$. Using instead the $1/[T(T-T_c)]^{\nu}$ from
  (\ref{3D-TIME}) gives $\nu \approx 0.67$. \label{3D-LENGHT}}
\end{figure}

We now present results of numerical simulations of
Hamiltonian~(\ref{3D}) on the three-dimensional cubic grid on the
Nishimori line. Studying the equilibrium dynamics on the Nishimori
line, using the results of section~\ref{Nishi}, is quite simple. We
simulate the melting process starting from the fully ordered
configuration, just as we did in the mean field system. Of course now
the system on the Nishimori line is not equivalent anymore to the spin
glass case with $\rho=1/2$, but this is irrelevant for our discussion.
If there is a first order phase transition on the Nishimori line, then
the melting process with a Vogel-Fulcher-Talman-like relaxation time
and diverging length scales implies that the equilibrium dynamics
undergoes an ideal glass transition described by the RFOT.

The correlation function for equilibrium dynamics (or equivalently the
magnetization in the melting process) is plotted versus temperature in
Fig.~\ref{3D-SIMU1}. It yields the standard picture with a growing
plateau and the melting time ---or equivalently the equilibrium
auto-correlation time--- grows faster than exponentially, and is
perfectly consistent with a VFT divergence at $T_c=0.88$, as attested
in Fig.~\ref{tau_3d}. Note that we use the usual form of the VFT, but
we could have used the generalized form eq.~(\ref{3D-TIME}) with
similar agreement but with slightly different values of $T_c$.

The upper part of Fig.~\ref{3D-SIMU1} shows the dynamical
susceptibility or equivalently the 4-point dynamical
susceptibility. Its maximum $\chi_4^{\rm max}$ is growing with the
growing relaxation time $\tau$. In \cite{US-PART-I} we observed for
the ferromagnetic two-dimensional non-disordered case that
$\chi_4^{\rm max}\approx \tau$. The present data for the $5$-spin
model on the Nishimori line instead indicate a low-exponent power law;
in fact for the largest time the best power-law fit gives $\chi_4^{\rm
  max}\approx \tau^{0.2}$. Such a low exponent is compatible with a
logarithmic growth in time; in fact a similar behavior is observed in
the activated coarsening of the random field Ising model and in the
random bond Ising model, where a logarithmic growth is expected
\cite{RFIM-th} but in numerical simulations a power-law with low
exponent is observed \cite{RFIM-num}. The low-exponent power law is in
this case usually attributed to pre-asymptotic effects. The slower
divergence of $\chi_4^{\rm max}$ in the presence of disorder (to be
compared with the non-disordered transition in the Potts model observed
in \cite{US-PART-I}) is thus due to an activated coarsening for the
growth of the nuclei, as expected.

We also investigated the point-to-set correlation function
\cite{PTS-Exp-Num}.  We consider a box of size $N=L^3$, apply fully
ferromagnetic boundary conditions, and measure the equilibrium
magnetization inside the box for different temperatures and
sizes. Again, since we are on the Nishimori line, this is nothing but
the point-to-set equilibrium correlation length, computed in a much
faster way. The data are plotted in Fig.~\ref{3D-PTS} and the
correlation length in Fig.~\ref{3D-LENGHT}. We observe a growing
length scale compatible with a power law divergence at $T_c$. By
fitting our data using the formula $\ell \propto [T (T-T_c)]^{-
  {\nu}}$ of eq.~(\ref{3D-nucl}) we found that $T_c$ is compatible
with the value $0.88$, and that the exponent $\nu$ is compatible with
the value $2/3$ advocated in \cite{KTW}. 

Given we barely span two decades in Fig.\ref{3D-LENGHT} one cannot be
truly conclusive. Still, all our numerical results for the $5$-spin
model on the 3-dimensional grid on the Nishimori line support the
existence of a first order transition close to $T_c \approx
0.9$. Since the melting process is equivalent to equilibrium dynamics
on the Nishimori line, this is a perfect realization of RFOT ideas in
a finite dimensional setting. Below the Kauzmann transition the
equilibrium physics is, however, given by the ferromagnetic phase, not
by an ideal glass phase. This is not a problem, as this is after all
what happens in real systems ---the crystal is always the correct
equilibrium state--- and as we are mainly interested in how a liquid
becomes a glass, not in the out-of-reach glass state below $T_K$
(which is anyway ill-defined \cite{STAR_model}).

\subsection{Is there really a first order transition?}
The numerical data are compatible with the existence of a first order
phase transition on the Nishimori line, and as a consequence, with the
existence of the ideal glass transition in a finite dimensional
system.  But is it really a genuine first order phase transition?  Of
course, this cannot be answered based only on numerical simulations,
and we cannot exclude that if one goes to much larger times and to
point-to-set lengths equal to, say, $\xi \approx 100$ a crossover to a
second-order transition arises. Or worse, there might be no phase
transition at all. Note, however, that such first-order transitions
definitely exist at the mean field level (see section \ref{mf-nish})
and that at least second-order ones do exist on the Nishimori line
(see \cite{Jesper}) in finite dimension: We are thus assured that a
transition can be found. At least, the issues and criticisms discussed
in \cite{Moore}, where it was argued that the RFOT transition in finite
dimension maps {\it only} to a spin glass droplet model in a field,
where it seems there is no transition at all \cite{Katz}, do not apply
in our case. The difficulty here is, however, to decide if on the
Nishimori line there can be a first order phase transition in finite
dimension.

Unfortunately, in all the solvable cases we found in the literature
the transition on the Nishimori line is of the second order,
e.g. \cite{Jesper}. This is not very surprising because all these
cases concern either 2-dimensional systems to which the theorem of
Aizenman-Wehr \cite{Self-3} about smoothening of first order phase
transition in the presence of disorder applies, either system which
have a 2nd order phase transition even in the absence of disorder.  If
a genuine first order transition can be found on the Nishimori line in
a 3D system is hence a challenging open question, which is clearly
worth investigating.

Proving the existence or impossibility of a first order phase
transition in a finite dimensional model on the Nishimori line would
be a nice achievement. A positive answer would prove that a true VFT
divergence exists in finite dimension, and that it is associated with
a RFOT mechanism. However, on the practical side, this is not such a
fundamental issue. If the transition becomes of 2nd order at sizes so
large that the equilibrium behavior cannot be observed in a time
smaller than any experimental time, then the first order approach for
the melting problem (and the RFOT one for the glassy dynamics) is the
correct description at the observable times scales.

\section{Discussion}
In this paper we have further studied the analogies between the
melting dynamics above a first order phase transition and the
equilibrium glassy dynamics discussed in \cite{US-PART-I} by exploring
a class of models where the two processes are strictly equivalent:
Ising spin systems on the Nishimori line (some disordered Potts models
share the properties discussed in this paper \cite{Jesper}). In the
light of the mapping between melting and equilibrium dynamics, we
investigated several features of the dynamics of super-cooled liquids:
the diverging relaxation time, the plateau in the correlation
function, dynamical heterogeneities related to the dynamical
susceptibility $\chi_4$, and the point-to-set correlations.  We show
that they all can be easily recovered, understood, and studied in the
much simpler, and more familiar, setting of first order phase
transition. Since the RFOT was invented in analogy with first order
transitions, it is rather interesting that there exist systems where
the two are equivalent.

In particular, on the Nishimori line, the dynamical ---or
mode-coupling--- transition corresponds to the spinodal point, while
the Kauzmann transition corresponds to the first order ferromagnetic
transition. The mosaic approach is replaced by simpler nucleation
arguments. We hope this approach will help in expanding our
comprehension of the glass transition problem. To conclude this work,
we would like to make several comments.

\paragraph*{\bf Ideal glass transition?} 
A very fundamental open question in the field of glassy systems is the
existence of a finite dimensional model where one can observe
the VFT divergence.  The bottom-line of our result here is that if
there is a first order phase transition on the Nishimori line then the
ideal glass transition exists. It would be very interesting to
prove (or disprove) the existence of such a transition on the
Nishimori line. Hopefully, this will be easier than the original
question, as on the Nishimori line we are discussing a simpler to
study first order phase transition. Moreover, the gauge transformation
can be used to prove many results rigorously \cite{Nishimori}.

\paragraph*{ \bf String-like motion} 
It is well known in the theory of nucleation and first order phase
transitions that the nucleating droplets can also be non-compact and
fractal \cite{ReviewBinder}. As noted by \cite{Wolynes}, 
the stringy nuclei observed close to the pseudo-spinodal \cite{KLEIN} in
 first-order transitions could be the analog of the string-like motion observed
 in the relaxation in glassy systems \cite{Sharon}. The results
 presented here and the exact analogy between melting and glasses thus
 provide a firmer motivation for this. It would also be interesting to
 observe the string-like motion during the melting process of
 superheated solids: This is maybe related to the {\it rings} and {\it
   loops} observed in bulk melting simulations \cite{MO}.

\paragraph*{\bf Mode coupling transition as a mean field approach} Many
theoretical results about glasses are based on the mode coupling
theory \cite{MCT,MCT2}. The MCT predictions are exact in the
mean-field $p$-spin model, and in this model the mode coupling
transition can be mapped to a spinodal point of a first order phase
transition on the Nishimori line. One hence cannot escape the
conclusion that the transition described by MCT is a kind of
mean-field melting, and the mode coupling temperature is a
corresponding spinodal point. Of course, this suggestion is present in
the literature since the work of \cite{KT,KW1,KW2}, but our mapping makes it
very explicit.

\paragraph*{\bf Simulating MCT behavior}
A straightforward but powerful application of our mapping is to
perform fast simulations of the equilibrium glassy dynamics in
mean-field models, as we have done for the $p$-spin, by simply
starting from the fully ordered configuration. This should allow us to
go beyond the current size limitations and investigate better
numerically the mean-field systems and the finite-size corrections
(in the spirit of e.g. \cite{SarlatBilloire09}). This is in fact what
we have started to do in section \ref{simu-super}.

\paragraph*{\bf Crossover MCT/finite dimension}
One of the crucial issues in the RFOT is to understand better
nucleations in glassy systems, and our mapping allows to put this
question back into the standard framework of first order phase
transitions. The crossover from the mode coupling behavior to the
finite dimensional activated dynamics (described e.g. by the mosaic
theory) then becomes the usual crossover from spinodal to activation
at a first order phase transition \cite{ReviewBinder}. A promising
direction in which these ideas could be extended is the Kac limit
\cite{SILVIO}, or the field theoretic computation of instanton in
order to estimate the free energetic barriers. Note also that the
system is guarantied to be replica symmetric on the Nishimori line
which may simplify many calculations.

\acknowledgments It is a pleasure to thank G. Biroli, J-P. Bouchaud,
A. Cavagna, S.~Franz, J. Kurchan, J. Langer and H. Yoshino for
interesting discussions and comments.

\end{document}